\documentstyle[12pt]{article}
\title{Method of effective potential for quantum
 Heisenberg ferromagnet theory}
\author{E.V.Podivilov\\
Institute of Automation \& Electrometry,\\
Novosibirsk, 630090, Russia}
\def\tr{\rm Tr}
\begin{document}
\maketitle
\begin{abstract}
Self - consistent temperature dependence of  the average
magnetization in quantum Heisenberg ferromagnet is
obtained as a first approximation of perturbation theory on an
inverse radius by application of the functional method for quantum
ferromagnet. The experimental data are compared with the theoretical
results for {\bf EuO} ferromagnet. A quantitative agreement is observed
between theoretical and experimental results outside interval near
Curie temperature. The theory is of critical behavior in the vicinity
of Curie temperature $T_c$. Theoretical $T_c$ value itself is agreed
with the experimental one with an accuracy of several percents.
\end{abstract}

\section{Introduction}

The functional representation of a partition function allows to
extend the application of the approximate methods to the problem of
the statistical physics.  Convenience and naturalness of the
functional representation resulted in numerous attempts to extend
this method to the Heisenberg quantum ferromagnet model (see
\cite{Leibler,Jevicki,Kolokolov,Popov}).  A method
proposed in \cite{Kolokolov} makes it possible to write an expression for the
partition function of a magnet in the asymmetric phase in the
explicit closed form of an integral over two number fields --- a
neutral field and a charged field. The charged field $\Psi$
corresponds to the elementary excitations of magnet. The neutral
field $\rho$ corresponds to the average spin magnitude.  Erroneous
interpretation of global effects in \cite{Kolokolov} was corrected in
\cite{Podivilov}.

The partition function
may be considered as a generating functional of averages of the spin
operators. Accordingly the closed expression for the partition
function allows to calculate approximately different averages of spin
operators, using small parameter \cite{Vaks} $1/z$ --- the inverse
range of the interaction.

A problem of the average spin calculation can be formulated within the
scope of the functional formalism as a minimum of the effective
potential $W_{eff}(\rho)$ problem \cite{Coleman}. We will calculate
$W_{eff}(\rho)$ in the first order in $1/z$. Minimizing
$W_{eff}(\rho)$, we will next obtain the average magnetization $M(T)$
as a function of the temperature.  Such a way of considering it
allows to avoid the emergence of the fictitious divergences presented
in the direct diagram expansion. The comparison between theoretical
relationship $M(T)$ and experimental one for the ferromagnet {\bf
EuO} demonstrates that they are in close agreement, just as magnitudes
of Curie's temperature with an accuracy of a few percents.

\section{Generating functional of averages of the spin operators}

The model of the Heisenberg quantum ferromagnet is based upon the
Hamiltonian

\begin{equation}
\hat H = - {1\over 2}\sum_{i,j}\hat{\vec S_i}J_{ij}\hat{\vec S_j},
\end{equation}
where $J_{ij}$ --- is the matrix of the exchange interaction between
neighboring spins $\hat{\vec S_i}$ which are located at the lattice
sites $\vec r_i$.  If the interaction of spins with external
magnetic field is taken into account the partition function can be
written as:

\[
Z[\vec h_i(t)] = \tr T \exp\left(- \beta \hat H +
\int\limits_0^{\beta} dt\, \vec h_i(t) \hat{\vec S_i} \right),
\]
where the symbol $T$ denotes time ordering and $\vec h_i(t)$ is the
external field at lattice site $\vec r_i$. For the operator
$\exp(-\epsilon \hat H)$ a Gaussian transformation

\[
\exp(-\epsilon \hat H) =\int\prod_i {\sqrt{\epsilon}d\phi_i \over
\sqrt{2\pi}} (\hbox{det}J_{ij})^{-1/2}
\exp \left(-\sum_{ij} \vec{\phi_i} J^{-1}_{ij}
\vec{\phi_j} \epsilon /2 + \vec{\phi_i} \hat{\vec S_i}\epsilon\right)
+ O(\epsilon^2)
\]
can be performed to within terms $\sim \epsilon^2$. Thus, writing
$\exp(-\beta \hat H) =$ $(\exp(-\epsilon \hat H))^{\beta /\epsilon}$,
with $\epsilon \to 0$, we arrive at an expression for the partition
function in the following form

\begin{equation}
Z[\vec h_i(t)] = N\int\prod_i D\phi_i(t)
\exp \left(-{1\over 2}\int_0^{\beta}dt\,
\sum_{ij} \vec{\phi_i}(t) J^{-1}_{ij} \vec{\phi_j}(t) \right)
\end{equation}
\begin{equation}\label{partition}
\times\prod_i\tr T \exp\left(\int\limits_0^{\beta} dt\,
[\vec{\phi_i}(t) + \vec h_i(t)] \hat{\vec S_i} \right),
\end{equation}
The substitution put forward by Kolokolov in \cite{Kolokolov}

\begin{eqnarray}
&\phi_i^z = J_{ij}\rho_j - h_i^z,\nonumber\\
&\phi_i^+ = \psi_i^+ - h_i^+,\label{substitution}\\
&\phi_i^- = {d\psi_i^- \over dt} - h_i^- - \psi_i^- J_{ij}\rho_j +
(\psi_i^-)^2 \psi_i^+,\nonumber
\end{eqnarray}
\[
\phi_j^{\pm} = {1\over 2} (\phi_j^x \pm i \phi_j^y)
\]
rearranges the ordered operator
exponential (\ref{partition}) to the explicitly specified
operator

\begin{equation}\label{Texp}
\hat A (\beta) = T \exp\left(\int\limits_0^{\beta} dt\,
[\vec{\phi}(t) + \vec h(t)] \hat{\vec S} \right) =
\exp (\psi^-(\beta) \hat S^+) \;
\end{equation}
\[
\times
\exp \left(\hat S^z \int_0^{\beta}dt\,\tilde{\rho}(t) \right)\;
\exp \left(\hat S^- \int_0^{\beta}\psi^+(t) \exp \left(\int_0^t
\tilde{\rho}(t') \, dt'\right) \, dt \right)\;
\]
\[
\times\exp (-\psi^-(0) \hat S^+),
\]
where index $i$ is omitted,

\[
\tilde{\rho}_i = J_{ij}\rho_j - 2 \psi_i^- \psi_i^+,\quad
\hat S^{\pm} = \hat S^x \pm i \hat S^y.
\]
Thus, regarding  $\psi^{\pm}(t)$ and $\rho(t)$ as new integration
variables, we can calculate the trace of the T-exponential
(\ref{Texp}) explicitly and obtain a closed functional representation
for $Z[\vec h]$. We have for the arbitrary spin magnitude $S$

\[
\tr \hat A (\beta) = \sum\limits_{l = 0}^{2S}
\left[\left(\psi^-(\beta) - \psi^-(0) \right)
\int\limits_0^{\beta} \, dt \psi^+(t)
\exp\left( \int\limits_0^t\tilde{\rho}(t') \, dt'\right)\right]^l
g_l\left(\int\limits_0^{\beta} \tilde{\rho}(t) \, dt\right),
\]
\[
g_0(\xi) = \sum\limits_{m = -S}^S e^{m\xi },\quad
g_l(\xi) = \sum\limits_{m = -S}^S P_1\times\dots\times P_l e^{m\xi},
\]
\[
\quad P_l = S(S+1) - (m+l)(m+l-1).
\]
The authors of \cite{Podivilov} showed that it is necessary to use the
initial conditions $\psi_i^-(0) = 0$ in order to transform from the
original variables to the new integration variables
(\ref{substitution}). With this conditions and in the discretization
of the transformation (\ref{substitution})

\begin{equation}\label{discretization}
{d\psi^-(t)\over dt} = \lim\limits_{\epsilon \to 0} {1\over \epsilon}
\left(\psi^-(t+\epsilon) - \psi^-(t)\right)
\end{equation}
the Jacobian of the transformation is equal to the constant.
Hence follows a functional representation for $Z[\vec h]$

\begin{eqnarray}\label{PF}
&Z[\vec h] = N\int\,\prod_i D\rho_i
D\psi_i^-D\psi_i^+\, \exp(-\Gamma_l -\Gamma_n), \\
&\Gamma_n = \sum\limits_i -\ln\left(\sum\limits_{l = 0}^{2S}
\left[\psi_i^-(\beta)
\int\limits_0^{\beta} \, dt \psi_i^+(t)
\exp\left( \int\limits_0^t\tilde{\rho}_i(t') \, dt'\right)\right]^l
g_l\left(\int\limits_0^{\beta} \tilde{\rho}_i(t) \, dt\right)\right)
\label{G_n}\\
&\Gamma_l = \int_0^{\beta} \, dt \sum\limits_{ij} {1\over 2}
\left[\rho_i J_{ij} \rho_j - 2\rho_i h_i^z +
h_i^z J_{ij}^{-1} h_j^z \right] +\nonumber\\
&2\left[ \left(\psi_i^+ - h_i^- \right) J_{ij}^{-1}
\left({d\psi_j^-\over dt} - h_j^+ -\psi_j^-J_{jk}\rho_k +
(\psi_j^-)^2 \psi_j^+\right)
\right].
\label{G_l}\end{eqnarray}
This partition function may be considered as a generating functional
of averages of the spin operators:
\[
 <\hat S_i^z(t)> = \lim\limits_{\vec h \to 0} {\delta \ln
 \left(Z[\vec h]\right)\over \delta h_i^z(t)} = <\rho_i(t)>.
\]

\section{Effective potential of field $\rho$}

In order to obtain the field $\rho$ effective potential , we must
integrate over fields $\psi^{\pm}$. We cannot perform this
integration exactly, however
it can be made approximately over parameter $1/z = \sum_j J_{ij}^2\;
/\; (\sum_j J_{ij})^2 = \sum_{\vec k} J^2_{\vec k}\;/\;J_0^2$, where
$J_{\vec k}$ is the Fourier transform of the exchange matrix $J_{ij}
= J(\vec r_i-\vec r_j)$ ($J_0 = J(\vec k = 0),\; \sum_{\vec k}
J_{\vec k} = 0$).

Let us expand the nonpolynomial part of
action $\Gamma_n$ (\ref{G_n}) in a series in powers of $\psi^-\psi^+$
and extract all the $\Gamma$ terms at $\vec h = 0$, which are linear
functions of $\psi^{\pm}$

\begin{eqnarray}\label{efpot}
&\Gamma^1 = \int\limits_0^{\beta} \, dt \sum\limits_{ij}
{1\over 2} \rho_i(t) J_{ij} \rho_j(t) + 2 \psi_i^+(t) J_{ij}^{-1}
 \left({d\psi_j^-\over dt} -\psi_j^-(t) J_{jk}\rho_k(t) \right)+
\nonumber\\
&\int\limits_0^{\beta} \, dt \sum\limits_i 2 b(\rho_i^0(\beta))
\left( \psi_i^+(t)\psi_i^-(t) - \psi_i^-(\beta) \psi_i^+(t)
\exp(\rho_i^0(t)) n_0(\rho_i^0(\beta))\right) -\nonumber\\
&\sum\limits_i g_0(\rho_i^0(\beta)),\\
&\rho_i^0(t) = \int\limits_0^t \sum\limits_j \,d\tau\, J_{ij}
\rho_j(\tau),
\quad n_0(\xi) = {1\over exp(\xi) - 1}, \nonumber
\end{eqnarray}
where $b(\xi) = d\ln(g_0(\xi))/d\xi$ is Brillouin function. The
contributions from the other terms of $\Gamma$ are small in the
inverse range of interaction, because these terms include two and
more sums of $(J_{ij})^2$.  Hence in a zero approximation on
$1/z$ follows:
\begin{enumerate}
\item
The saddle-point value $\rho$ is determined by equation

\begin{equation}\label{S}
<S^z> \simeq {\overline \rho}^0 = b (\beta J_0 {\overline \rho}^0),
\end{equation}
which agree with mean field theory equation.

\item
The Fourier transform of the longitudinal correlator
$K_{\vec k}$ takes the form

\begin{equation}\label{K}
K_{\vec k}(t,t') = <(\rho_{\vec k}(t) - {\overline \rho})
(\rho_{\vec k}(t')- {\overline \rho})> - J_{\vec k}^{-1}\delta(t-t')
\simeq {b'(\beta J_0 {\overline \rho})\over 1 - \beta J_{\vec k}
b'(\beta J_0 {\overline \rho}) },
\end{equation}
where ${\overline \rho} = {1\over M}\sum\limits_i \rho_i = \rho(\vec
k = 0)$ is the average value of the field $\rho_i$, $M$ --- number of
lattice sites.

\item The bare propagator of the field $\psi^{\pm}$  in the Fourier
representation has the form

\begin{equation}\label{Gr}
G_{\vec k}(t,t') = <\psi^-_{\vec k}(t) \psi^+_{\vec k}(t')> \simeq
{J_{\vec k}\over 2}\left(\theta (t-t') + (n_{\vec k} + 1)
\left[\exp(J_{\vec k} b t) - 1\right] \right)\times
\end{equation}
\[
\exp(\omega_{\vec k}(t-t')),
\]
\[
n_{\vec k} = {1\over \exp(\omega_{\vec k}\beta) - 1},\quad
\omega_{\vec k} = J_0{\overline \rho} - J_{\vec k} b,\quad
b = b(\beta J_0 {\overline \rho}),\quad \theta(0) = 1,
\]
where $ \theta(t)$  --- theta function.
\item The transverse correlator in the Fourier
representation has the following form

\[
K_{\vec k}^{-+}(t,t') = 4J_{\vec k}^{-2}<\phi^-_{\vec k}(t)
\phi^+_{\vec k}(t')> - 2J_{\vec k}^{-1}\delta(t-t') \simeq
\]
\begin{equation}\label{K-+}
 2b\left(n_{\vec k} + 1 - \theta (t-t')\right)
\exp(\omega_{\vec k}(t-t')).
\end{equation}
\end{enumerate}

Let us remind that we impose the zero initial conditions $\psi^-(0)
= 0$ on the $\psi^-(t)$ field, then the propagator is of the
$G_{\vec k}(0,t') = 0$ property and is nonperiodical. Function
$\omega_{\vec k}$ does not approach zero value at $\vec k = 0$ for an
arbitrary value of $\overline \rho$. However, if $\overline \rho$ is
defined by (\ref{S}) in the zero approximation on $1/z$, the
Goldstone symmetry will be restored ($\omega(0) = 0$).

By substituting $\Gamma^1$  for $\Gamma_n + \Gamma_l$ and
integrating (\ref{PF}) over $\psi^{\pm}$, we obtain the partition
function in the first approximation on $1/z$.

\begin{equation}
 Z = N\int\,\prod_i D\rho_i\, \exp(-W^1(\rho_i)),
\end{equation}
where up to a constant $W^1(\rho_i)$  has the form $\Gamma^1$ in
which the bare propagator (\ref{Gr})  is substituted for
$\psi^-(t)\psi^+(t')$. It can be pointed out that $W^1(\rho_i)$
also determines a longitudinal correlator, whose infrared behavior
was examined in low-temperature limit in \cite{Podivilov}.

Let us
perform the Fourier transformation of field $\rho_i$ and expand $W^1$
into a series in $\rho(\vec k \neq 0)$. It is enough to hold the
quadratic terms in $\rho(\vec k \neq 0)$ only within the first order
in $1/z$.  By integrating over $\rho(\vec k \not= 0)$, we obtain the
effective potential $W^1({\overline \rho})$.  The saddle-point value
$\overline\rho$ is determined in the first approximation on $1/z$ by
minimizing the effective potential $W^1({\overline \rho})$:

\begin{equation}\label{S1}
{\overline\rho} = b - \sum\limits_{\vec k} (1 - \beta J_{\vec k} b')
(n_{\vec k} - n_0) + {b''\over 2}\sum\limits_{\vec k} {\beta
J_{\vec k}\over 1 - \beta J_{\vec k} b'},
\end{equation}
This equation was obtained in \cite{Vaks} by spin diagram technique
application, wherein (\ref{S1}) was concerned as the approximation to
(\ref{S}). Within the scope of our approach the equation (\ref{S1})
means the equation in the saddle-point value $\overline\rho$ in the
effective potential, so there is a need to solve (\ref{S1}) exactly
for calculation $<S^z> = \overline\rho$. The temperature dependence
of the $S^z(T)$ at $S^z \to 0$ can be found analytically.
Magnetization $S(T)$ approaches zero as a square

\begin{equation}\label{crit}
S(T) = A \left(T^1_c - T\right)^{-{1\over 2}}
\end{equation}
at temperature $T^1_c$
\begin{equation}
T^1_c = T^0_c \left[
1 - {1 \over 12 b'(0)} \left({T^0_c\over T^1_c}\right)^2
\sum\limits_{\vec k} {J_{\vec k}^2 \over J_0^2} +
{5 b'''(0)\over 6 b'(0)} \sum\limits_{\vec k} {J_{\vec k} \over
T^1_c - b'(0) J_{\vec k} } \right],
\end{equation}
\[
T^0_c = b'(0) J_0 = {S(S+1)\over 3} J_0,
\]
which is the critical temperature value in the first approximation on
$1/z$.

For the illustration of the results obtained, let us compare
magnetization-temperature relationship (\ref{S1}) with that obtained
in experiment. We have pick Heisenberg ferromagnet {\bf EuO} as a
standard for comparison. Its structure (FCC --- face-centered cubic
lattice), exchange integral values for interaction between
neighboring spins $J1 = 1.212^0 K$ (12 neighbors) and $J2 = 0.238^0
K$ (6 neighbors), magnetization-temperature relationship $S(T)$ and
Curie's temperature $T_c = 69.15^0 K$ were examined carefully in
experiments \cite{Passell}. In {\bf EuO} the magnitude of spin is equal
to $S = 7/2$ and the perturbation theory parameter is $ \sum_{\vec k}
J^2_{\vec k}\;/\;J_0^2 = 1/14,\;J_0 = 15.97^0 K$. The figure
illustrates experimental relation between temperature and
magnetization $S(T)$ (points) and relation given by mean field theory
equation (\ref{S}) (curve 2) and given by equation (\ref{S1}) (curve
3) which were calculated by computer. Similar comparison between
theoretical results obtained by spin diagram technique and
experimental data on {\bf EuO} and {\bf EuS} were made in \cite{euo}.
The jump of curve 3 at $65^0$ K determines critical region. The
method of steepest descent with the help of which the equation
(\ref{S1}) was obtained, does not work in the critical region. It
means that dependence $S(T)$ specified by (\ref{crit}) breaks down
over critical region in the vicinity of $T_c$. However, $T_c$ itself
can be obtained exactly, since the effective potential dependence
$W(\rho)$ changes qualitatively at $T = T_c$. For $T < T_c$ function
$W(\rho)$ has two minimums at $\rho = \pm {\overline\rho}$, and for
$T > T_c$ function $W(\rho)$ has one minimum at $\rho = 0$.  Hence,
calculating effective potential at each following order of
perturbation theory on $1/z$, we can obtain more exact equation in
magnetization as well as critical temperature value.

It is significant that the critical behavior which is characteristic
to the phase transitions of the second kind in the method of
effective potential is conserved in the perturbation theory.
Therefore, in spite of inapplicability of our approach near $T_c$,
it nevertheless does not disturb the qualitative picture of the
phenomenon.

The author is thankful to I.V. Kolokolov for numerous discussion and
a help, and also to V.S. L'vov for interest to the work.

\end{document}